\documentclass{vgtc}                          

\ifpdf
  \pdfoutput=1\relax                   
  \usepackage{graphicx}                
  \DeclareGraphicsExtensions{.pdf,.png,.jpg,.jpeg} 
\else
  \usepackage{graphicx}                
  \DeclareGraphicsExtensions{.eps}     
\fi%

\graphicspath{{figures/}{pictures/}{images/}{./}} 

\usepackage{microtype}                 
\PassOptionsToPackage{warn}{textcomp}  
\usepackage{textcomp}                  
\usepackage{mathptmx}                  
\usepackage{times}                     
\usepackage{cite}                      
\usepackage{tabu}                      
\usepackage{booktabs}                  

\onlineid{0}

\vgtccategory{Research}

\vgtcinsertpkg

\usepackage{tikz}
\usepackage{xcolor}
\definecolor{FhG}{HTML}{179C7D}

\newcommand*\annotatedFigureBox[4]{%
    \draw[FhG,very thick,rounded corners] (#1) rectangle (#2);\node at (#4)%
    [fill=white,very thick,shape=circle,draw=FhG,inner sep=2pt,font=\sffamily,text=black]%
    {\textbf{#3}};%
}

\newenvironment{annotatedFigure}[1]{%
    \centering%
    \begin{tikzpicture}%
        \node[anchor=south west,inner sep=0] (image) at (0,0) {#1};%
        \begin{scope}[x={(image.south east)},y={(image.north west)}]%
    }{\end{scope}\end{tikzpicture}%
}

\newcommand{\product}{\emph{medhub}}
\newcommand{\productc}{\textcolor{FhG}{\product}}

\usepackage{enumitem}
\usepackage{todonotes}

\title{Towards \product: A Self-Service Platform for Analysts and Physicians}

\author{%
    Markus Höhn\thanks{e-mail: markus.hoehn@igd.fraunhofer.de}\\ %
    \scriptsize Fraunhofer IGD, Darmstadt %
    \and %
    Hendrik Lücke-Tieke\thanks{e-mail: hendrik.luecke-tieke@igd.fraunhofer.de} \\ %
    \scriptsize Fraunhofer IGD, Darmstadt %
    \and %
    Jan Burmeister\thanks{e-mail: jan.burmeister@igd.fraunhofer.de}\\ %
    \scriptsize Fraunhofer IGD, Darmstadt %
    \and %
    Jörn Kohlhammer\thanks{e-mail: joern.kohlhammer@igd.fraunhofer.de}\\ %
    \parbox{1.4in}{\scriptsize \centering Fraunhofer IGD, Darmstadt \\ TU Darmstadt}}

\teaser{
    \begin{annotatedFigure}
	    {\includegraphics[width=\linewidth]{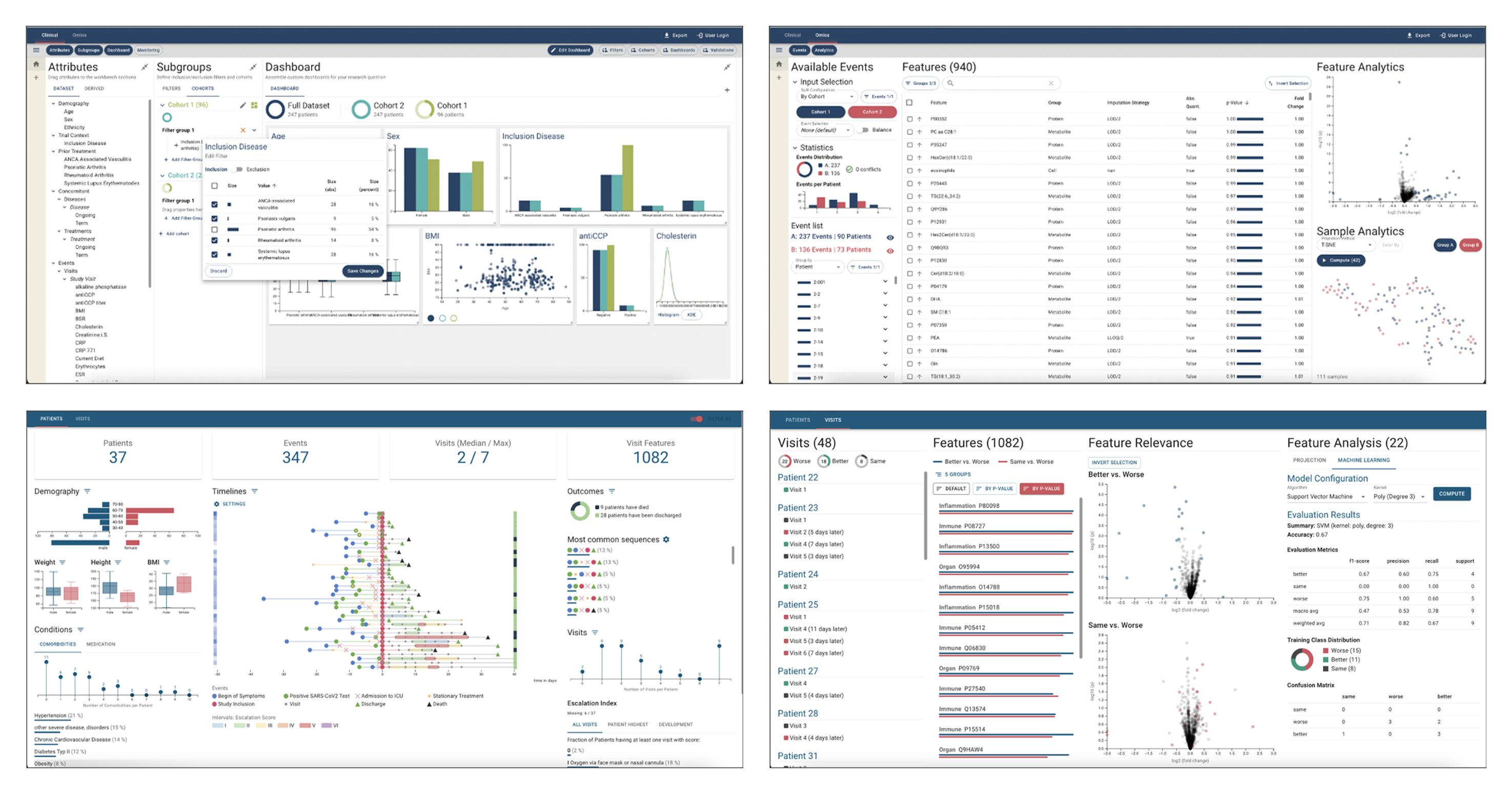}}
	    \annotatedFigureBox{0.000, 0.500} {0.500, 1.000} {A} {0.026, 0.950}
	    \annotatedFigureBox{0.500, 0.500} {1.000, 1.000} {B} {0.526, 0.950}
	    \annotatedFigureBox{0.000, 0.000} {0.500, 0.500} {C} {0.026, 0.450}
	    \annotatedFigureBox{0.500, 0.000} {1.000, 0.500} {D} {0.526, 0.450}
    \end{annotatedFigure}
    \caption{
        Upper row $\vert$ Data analysis workflow: (A) Creating individual cohorts with the help of a tree view containing all available attributes.
        The analyst can arrange dashboards to gain insights of the current cohorts to archive a sophisticated separation of the data.
        (B) A detailed overview with statistics, omics feature analytics and sample analytics based on the given cohorts provides the basis to discover new pathways and interdependencies between single features.
        Lower row $\vert$ Physicians' workflow: (C) Brushing and linking across multiple panels allows the physicians to analyze predefined cohorts - mostly healthy vs ill subjects.
        (D) A dashboard gives physicians more information on the omics-features concerning the given cohorts.
    }
    \label{fig:teaser}
}

\abstract{
    Combining clinical and omics data can improve both daily clinical routines and research to gain more insights into complex medical procedures.
    We present the results of our first phase in a multi-year collaboration with analysts and physicians aiming at improved inter-disciplinary biomarker identification.
    We also outline our user-centered approach along its challenges, describe the intermediate technical artifacts that serve as a basis for summative and formative 
    evaluation for the second project phase. 
    Finally, we sketch the road ahead and how we intend to combine visualization research with user-centered design through problem-based prioritization. 
} 

\CCScatlist{
  \CCScatTwelve{Human-centered computing}{Visualization}
  {Visual analytics}{};
  \CCScatTwelve{Information systems}{Information systems and applications}{Decision support systems}
  {};
  \CCScatTwelve{Applied computing}{Life and medical sciences}{}{}
}

\begin{document}

\maketitle

\section{Introduction} 
Finding less invasive methods to improve diagnosis and prediction of diseases are an important goal of biomarker identification \cite{brase2010serum}. 
Translating these research insights into clinical practice has the potential to improve health care, either by improving quality, swiftness or cost of prediction, diagnosis, and treatments.
However, every study conducted, every disease investigated, every machine used, and every stakeholder in this process of identifying and translating research into medical practice has its own set of requirements and constraints.

Gathering all requirements and engineering a generic solution that fits all requirements seems infeasible.
Furthermore, analysts and physicians alike emphasized the importance of solutions that fit into their individual workflows and also facilitate interdisciplinary communication and analysis. 
The different stakeholders have conflicting requirements that need to be satisfied if the overarching goal of quick turn-around times should be achieved. 
We approach this challenge by providing a flexible self-service platform called \productc.

\section{Related Work}
Aigner \& Miksch showed that the integration of classical data visualization and visualization of treatment data is an important source to find reasons and explanations to certain phenomena~\cite{aigner2006CareVis}.
They focused on user-centered design to develop an approach that was guided by physicians and medical experts.
%
Burmeister et al. demonstrated that self-service tools are a viable option for analysts to improve efficiency, and cover data preprocessing, cohort building, exploration, as well as reporting \cite{burmeisterSelfServiceDataPreprocessing2019}. 
However, they focus on medical researchers, which combine the role of analysts and physicians, while in our case these roles are distinct and the analysts regularly need to discuss with physicians about the matter at hand.
This is due to the complexity of our data with more than 1,500 attributes per patient observation which easily occludes facts that physicians can explain through their experience.
%
Metabolic pathway analysis has been widely used to detect features associated with complex disease phenotypes.
Maghsoudi et al. have shown that a multi-omics setup allows for multiple perspectives on the same problem to discover novel pathways and biomarkers~\cite{maghsoudi2022comprehensive}.
%
All these approaches provide a good basis for \productc, but offer only static visualization environments for patient and omics data used by analysts and physicians.

\begin{figure}[t]
    \centering
    \includegraphics[width=\linewidth]{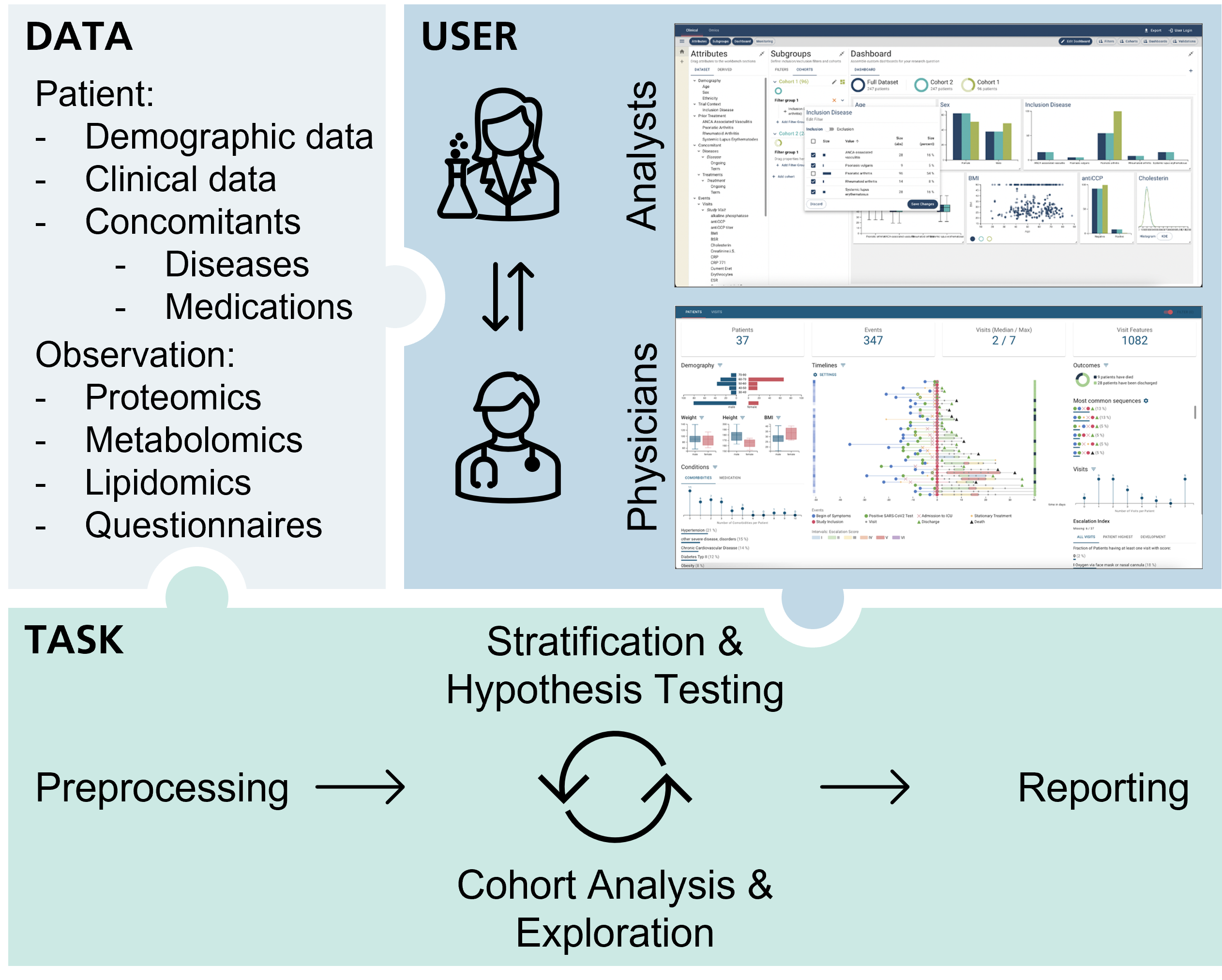}
    \caption{
        The concept of our approach uses the data-user-task triangle by Miksch et al. \cite{miksch2014matter}.
        Since there are two user groups with different usage scenarios, the analysts and physicians are obligated to communicate.
        This demand is also reflected in the task description.
    }
    \label{fig:concept}
\end{figure}

\section{Approach}
While a standardized workflow that enables fast and effective biomarker discovery has been proposed recently \cite{rischkeSmallMoleculeBiomarker2023}, the processes to translate these insights into medical practice are still evolving. 
One reason for this is the variety of data and workflows. 
Integrating drugs, devices, data and diagnostics into a single workflow is a promising step into the direction of improving the quality as well as the quantity of such findings \cite{phillips_diagnostics_2006}. 
Stakeholders involved in this process are analysts who do quantitative analysis to identify potential biomarkers on the one hand, and physicians who use those biomarkers to assess the health status of patients on the other hand. 
Bridging this gap is the main problem we want to solve with our research.
Since physicians have very limited time, the efficiency of discussions between stakeholders has to be optimized, which is our primary driver to prefer dynamic dashboard-building over static reports. 
Maximizing the efficiency for a given time frame offers a lot of potential for improved collaboration.
Physicians have extensive experience and knowledge of the specific diseases and their specific characteristics. 
This domain knowledge is highly valuable for analysts to define appropriate research goals.
Analysts might have a more technical background without much domain knowledge.
Yet, physicians can benefit from the knowledge gained from the data analysis to improve their treatments. 
For example, the observation that a biomarker is more clearly detectable in one tissue than in another may lead to new workflows in clinical practice.

%
%
%
\paragraph{Usage Scenario}
As shown in \autoref{fig:concept} we adapt the data-user-task approach of Miksch et al.~\cite{miksch2014matter} as our concept to support this interdisciplinary exchange.
Our main approach is a flexible user interface to configure individual dashboards to communicate between the analysts and physicians.
Analyst prepares a dashboard to discuss relevant questions with the physician.
Then, in the meeting, the dashboard can be expanded on the fly so that upcoming questions can be answered.
Hence, the efficiency of meetings can be improved.

\paragraph{Workflow}
The user is given a variety of well-known visualizations like bar charts, line charts, histograms for demographic attributes like age or gender, and clinical data like longitudinal data or a list of subscribed medications.
The user can compare two attributes directly by using scatter plots
to visualize the underlying measurements.
All the dashboards can be combined with individual filters and further user settings.
The user can drag and drop features from the given data set to a dedicated cohort-building area.

\paragraph{Benefit}
With this individual data preprocessing users can define cohorts interactively in an easy way.
The predefined dashboards provide a good insight into the data, whereas the predefined filters can be applied to the cohorts to refine them and the results can be shared and discussed with different stakeholders.
With a self-service platform, it is easy for users to extend and change the content of a dashboard on collaborators' demand, so that ad-hoc questions can be answered quickly and without resorting to a follow-up meeting. 






\begin{figure}
    \centering
    \includegraphics[width=\linewidth, trim=150 0 150 0, clip]{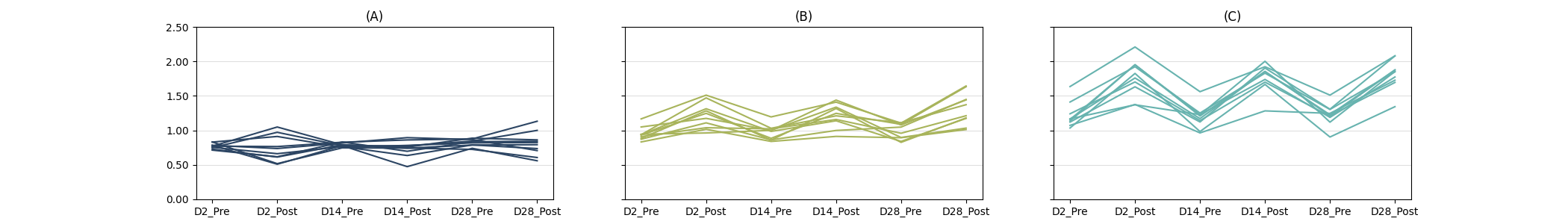}
    \includegraphics[width=\linewidth]{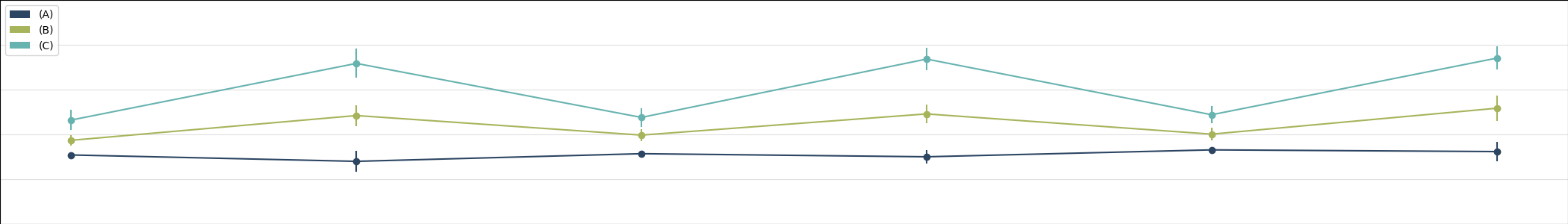}
    \includegraphics[width=\linewidth]{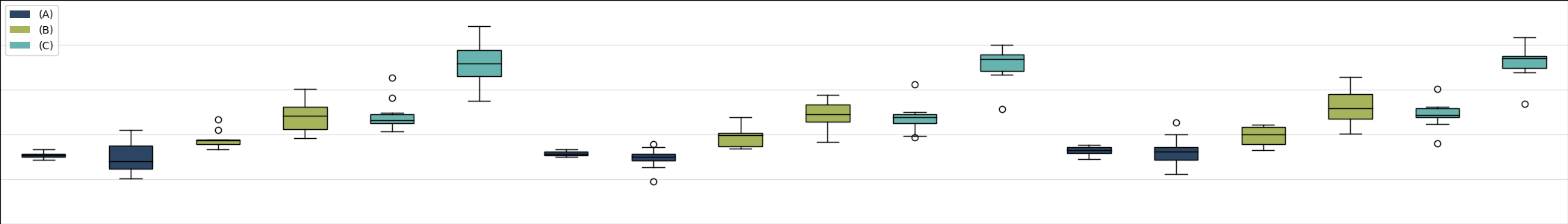}
    \caption{
        Data consist of three cohorts (A), (B), and (C), each with ten patients $\times$ six time points. 
        The top row shows multi-line plots for each cohort.
        The middle row shows a line for each cohort aggregated for each time point and enriched with a 95\% confidence interval.
        The bottom row shows the same data aggregated into box plots to visualize the distribution per cohort at the given time points.
    }
    \label{fig:fw}
\end{figure}

\section{Discussion \& Future Work}

To get more insights on individual omics features it is already possible to visualize cohorts by multi-line charts per cohort, aggregated data with confidence interval and box plots (see \autoref{fig:fw}).
In our discussions with domain experts, it became clear, that they are eager to try our current tools themselves.
Hence, we currently prepare workshops that assess feedback on the current state of our tools
and gather formative feedback on how to optimize for an interdisciplinary collaboration.

Once the base application has reached maturity to be deployed on-site, we will prioritize our next steps based on the first round of feedback.
In any case, we will cover additions that provide solutions for problems already brought up by our collaborating analysts and physicians. 
Current ideas include: 
(1) integrating visualization recommendations into the dashboard building to guide users to interesting correlations, 
(2) including pathway information to detect functions associated with complex disease phenotypes, and
(3) spatialization methods to generate visual cohort fingerprints or immunological landscapes to improve sensemaking over complex relationships between cohorts. 



\bibliographystyle{abbrv-doi-hyperref-narrow}

\bibliography{biblio}
\end{document}